\documentclass[12pt,a4paper]{article}
\usepackage{amsmath}
\usepackage{graphicx}
\usepackage{textcomp} 
\usepackage{subfigure}
\usepackage{chicago}
\usepackage{color}

\begin{document}

\title{\vspace*{-22mm}\bf Competing metabolic strategies in a multilevel selection model}
\author{Andr\'e Amado$^{\dag}$, Lenin Fern\'andez$^{\dag}$, Weini Huang$^{\ddag}$ \\ Fernando F. Ferreira$^{\S}$, Paulo R. A. Campos$^{\dag}$}

\maketitle

\noindent
$^{\dag}$Evolutionary Dynamics Lab, Department of Physics, Federal University of Pernambuco, 50670-901, Recife-PE, Brazil \\
$^{\ddag}$Department of Evolutionary Theory, Max Planck Institute for Evolutionary Biology, August-Thienemann-Stra\ss e 2, 24306 Pl\"on, Germany \\
$^{\S}$Center for Interdisciplinary Research on Complex Systems, University of S\~ao Paulo,
03828-000 S\~ao Paulo, Brazil 

\bigskip

\noindent Corresponding author: Paulo R. A. Campos 

\bigskip

\noindent Running Title: Competing metabolic strategies

\bigskip

\noindent e-mail: paulo.acampos@ufpe.br

\bigskip


\bigskip
\bigskip

\newpage

\begin{abstract}
The interplay between energy efficiency and evolutionary mechanisms is addressed. 
One important question is the understanding of how  evolutionary mechanisms can select for the optimised usage
of energy in situations where it does not lead to immediate advantage. 
For example, this problem is of great importance to improve our understanding about the major transition from unicellular to 
multicellular form of life. The immediate advantage of gathering
efficient individuals in an energetic context is not clear.
Although this process tends to increase relatedness among individuals, it also increases local competition. 
To address this question, we propose a model of two competing metabolic strategies that makes
explicit reference to the resource usage. We assume the existence of 
an efficient strain, which converts resource into energy at high efficiency but displays a low rate of 
resource consumption, and an inefficient strain, which consumes resource at a high 
rate with a low efficiency in converting it to energy.
We explore the dynamics in both well-mixed and structured populations.  
The selection for optimised energy usage is measured by the likelihood of
that an efficient strain can invade a population only comprised by inefficient strains. 
It is found that the region of the parameter space at which the efficient strain can thrive in structured populations is always larger 
than observed in well-mixed populations. In fact, in well-mixed populations the efficient strain is only evolutionarily stable
in the domain whereupon there is no evolutionary dilemma. We also observe that small group sizes enhance the chance of invasion
by the efficient strain in spite of increasing the likelihood of competition among relatives. This outcome corroborates 
the key role played by kin selection and shows that the group dynamics relied on group expansion, overlapping generations and group 
split can balance the negative effects of local competition.
\end{abstract}
\newpage

\section{Introduction}
\label{sec:introduction}
The handling of a common resource by individuals with different means to exploit the resource often gives rise to social conflicts \shortcite{HardinScience1968}. One of the well-known examples is observed when these features are related to how rapidly and
how efficiently the resource is exploited. With the massive available data from experiments, especially in 
microbial populations, the social conflict 
ensued by resource competition has been more effectively addressed \shortcite{MacleanHeredity2008}.
The social conflict arises directly from the trade-off between growth rate and yield of resource exploitation.
Empirical studies demonstrate that the trade-off between resource uptake and yield is commonplace in the
microbial world \shortcite{MeyerNatComm2015,KapplerMicrobiology1997,PfeifferScience2001,OtterstedtEMBO2004}. The trade-off owes 
to biophysical limitations which prevent organisms from optimizing 
multiple traits simultaneously. Independent experiments have reported the existence of a negative correlation between rate and
yield \shortcite{MeyerNatComm2015,NovakAmNat2006,LipsonBiogeo2009,PostmaApEnvMic1989}. The existence of the trade-off for 
ATP-producing reactions can be derived using arguments from irreversible thermodynamics \shortcite{MacleanHeredity2008,HeinrichEurJBio1997}. 
\bigskip

Important insights have been gained regarding the understanding of energy efficiency at the cell standpoint. 
By energetic efficiency we mean the ability of cells to extract energy for a given amount of resource. It was
shown that gene expression across the whole genome typically changes with growth rate \shortcite{WeibePNAS2015}. This 
evidence naturally emerges from the 
assumption of occurrence of different trade-offs at the intracellular level \shortcite{WeibePNAS2015,MaitraPNAS2015}, regardless of the metabolic pathway used to convert resource into ATP. 
The pathway choice dictates the speed and efficiency of growth. 
Heterotrophic organisms are good examples of the yield-rate trade-off. Their metabolism converts 
resource like glucose into energy in the form of ATP, mainly through two processes: fermentation and 
respiration \shortcite{RichBST2003,HellingJBac2002}. Fermentation does not require oxygen and uses glucose as a reactant 
to produce ATP. However, it is considered to be inefficient because the end products still contain a great deal of chemical energy.
On the other hand, respiration produces energy from glucose after it is released from its chemical bonds due to a molecular breakdown.
It is efficient as several ATPs are produced from a single glucose molecule \cite{PfeifferTrends2005}. 
A microorganism using fermentation depletes the resource faster than microorganisms using respiration. 
However, respiration allows cells to make a better use of resource and to produce more offspring given the same amount of resource.
In a population of cells using different metabolic processes, there will be competition between high efficient cells (respiration) and 
inefficient cells (fermentation). 

\bigskip

This competition among traits presenting distinct metabolic pathways is the concern within a broad
evolutionary perspective. The physiological states of a cell play an important role. At the physiological level, previous 
investigations have tried to understand the underlying conditions that may trigger the switch from one mode of metabolism
to another \shortcite{OtterstedtEMBO2004}. The most influential factor studied so far is the resource influx rate into
the system.  This problem has been addressed in the framework 
of spatially homogeneous environments \shortcite{FrickNatur2003} as well as in spatially structured environments 
of unicellular organisms \shortcite{PfeifferScience2001,AledoJMolEvol2007}. The studies of the steady state 
of the population dynamics were carried out either by writing down a set
of coupled differential equations describing population density and resource dynamics or 
by directly  mapping the problem into a prisoner-dilemma game, thus requiring some arbitrary 
definition of a payoff matrix in terms of the resource influx rate. 
Conclusions can then be drawn from
the established results in the field of evolutionary game theory. 
From those studies, it is well established that in homogeneous environments 
competition always drives the strain with the lowest rate of ATP production 
to extinction \shortcite{PfeifferScience2001}. On the other hand, when space is structured
 the outcome of the system is determined by its free energy content (resource influx rate). If the resource is abundant, the strain using 
the pathway with high rate and low yield dominates. If the resource is limited, the strain using the pathway with high yield but at low growth rate 
dominates \shortcite{PfeifferScience2001}. 

\bigskip

Despite the attempts to study this trade-off on structured populations, these works have focused on 
spatially structured populations \cite{PfeifferScience2001,AledoJMolEvol2007}. 
While this provides a valuable 
approach to study biological structures like biofilms, it lacks a clear definition of group and group 
reproduction, essential features of multicellular organisms as they are usually defined in animals, plants and 
fungi \shortcite{claessen2014bacterial}. In this sense these models describe more a
 multi-organism community rather than a multicellular organism. It is known that multicellular life 
can appear without a biofilm-like phase, for example through the experimental evolution of 
multicellularity \shortcite{RatcliffPNAS2012}. Our approach does not depend explicitly on spatial structure, instead we focus
on the group structure. This way, we are able to study multicellular structures that can reproduce as groups. 
It is also critical to understand which evolutionary force can turn these multicellular structures into evolutionary stable units. 

\bigskip

One potential driving force promoting the evolution of the cooperative strain is kin 
selection \shortcite{Hamilton1964}. The theory of inclusive fitness states that  
individuals have evolved to favor those genetically related to themselves. The fundamental principle of kin selection is
formalized in Hamilton's rule --- a cooperative trait can increase in frequency if $rb>c$, where $r$ is a measurement of genetic
similarity between the recipient and the donor of the cooperative act, $b$ is the 
benefit received by the recipient, and $c$ is the cost of the trait to the donor's  fitness \shortcite{Hamilton1964,WestScience2002}.  Therefore, a certain level of genetic relatedness
between individuals is required for cooperation to evolve \shortcite{WestCurrBiol2007}, and it can be achieved, for example, by 
kin discrimination. While being 
common among social organisms from microbes to humans  \shortcite{RussellPRSLB2001,RenduelesPNAS2015},
it requires complicated cognitive faculties to allow organisms to discriminate 
between kin and non-kin \cite{RoussetEvolution2007}. Another key mechanism of  the high relatedness is limited dispersal
owing to population viscosity \shortcite{Hamilton1964,RodriguesEvolution2012,RodriguesEvolution2013}, which means that offspring disperse slowly from the sites of their origin \shortcite{DayTREE2011}.
However, although limited dispersal increases relatedness between interacting individuals, the benefit of kin selection
can be reduced or even mitigated due to local competition for resources among the 
relatives \shortcite{WestScience2002,TaylorEvolEco1992,WestCurrBiol2007}. There 
are theoretical \shortcite{LehmannPhilRoyal2010,RodriguesEvolution2012,RodriguesEvolution2013} and experimental 
studies \shortcite{GriffinNature2004,KummerliEvolution2009} about the mechanisms 
to reduce the detrimental effect of local competition among kin. 
Limited dispersal has also been suggested as an explanation for the cooperative 
strategies in the evolution of metabolic pathways \shortcite{PfeifferScience2001,PfeifferPNAS2003}. However, there 
is no ultimate answer for how it promotes cooperation, as its effect will largely depend upon biological 
details \shortcite{GardnerJEvolBiol2006,MoudenJEvolBiol2008,LehmannEvolution2006}.

\bigskip
The current work aims to investigate the conditions under which the optimized handling of 
resource can prevail when selection acts
at different levels of biological organization. As such, a model of structured populations with structured groups is presented. 
The model assumes explicit competition among individuals for resource.  Although resource competition occurs 
at the cell level, the groups are assumed to replicate and spread 
at a rate which depends on its composition, putting the approach within the framework of multilevel selection.
Therefore, a higher level of organization is formed  and the group itself 
is seen as a replicating entity, a premise for the establishment of a multicellular organism. 
The issue especially motivated as multicellularity 
is conjectured to have arisen almost concomitantly with
the great oxygenation event, and thereby allowing the evolution of the respiration mode of 
cellular metabolism. Here, the problem is mainly addressed within an stochastic context whereby an evolutionary invasion analysis
is considered. We make use of extensive computer simulations, though analytical developments are also carried out for 
well-mixed populations.
A great limitation of studying population dynamics by means of deterministic models \shortcite{PfeifferScience2001} is 
that stochastic effects are neglected, and those
are particularly important in small populations, a very likely situation if the resource is not abundant. 
On the other hand, the problem has been mapped into a prisoner-dilemma game with constant payoffs \shortcite{FrickNatur2003,AledoJMolEvol2007},  which leads to a simplified assumption
that interactions among individuals are linear. Our approach avoids these limitations and 
allows us to make a thorough exploration of the cellular properties and
measure their effects on the population dynamics. This is certainly a gap
in the current literature, as pointed out by MacLean \shortcite{MacleanHeredity2008}.

\bigskip
The paper is organised as follows. In Section \ref{sec:material}, we present our model and the simulation methods. In Section \ref{sec:results}, we derive the equilibrium solutions, perform an invasion analysis, and explain the simulation results. In Section \ref{sec:conclusions}, we give a detailed discussion of our results.

\section{Methods}
\label{sec:material}
\subsection{Resource-based model description}
\label{subsec:model_description}

We consider a structured population with groups.
The population consists of individuals which use either of the two competing metabolic strategies: (1) efficient 
use of resource, denoted by $C$,  at which  an individual converts resource into ATP efficiently and thereby 
achieves a high yield, and (2) rapid metabolism, hereby denoted by $D$, with which an individual consumes 
resource at a high rate but converts it into ATP inefficiently. 

\begin{figure}[htp]
\centering
\includegraphics[width=\textwidth]{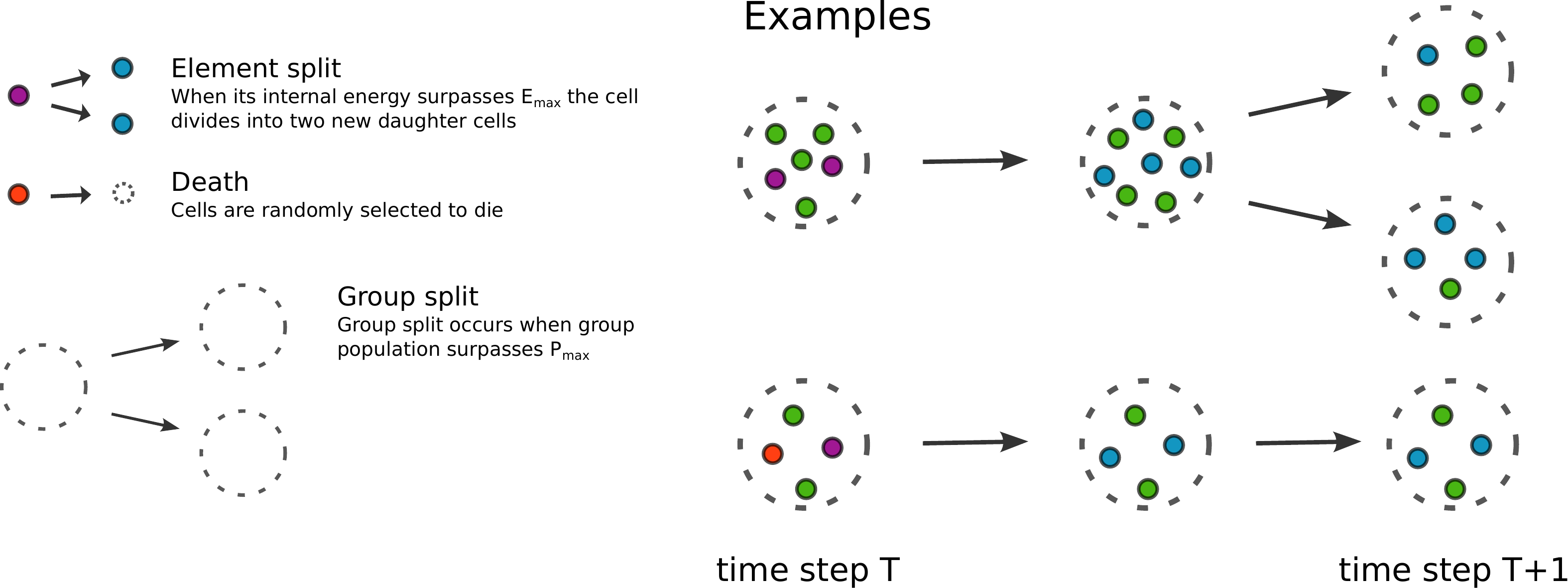}
\caption{A graphical representation of the processes of group splitting, cell division and death.}
\label{}
\end{figure}

\bigskip
The number of groups $N_G$ and their local 
population sizes $P_i$ ($i=1,..., N_G$) vary over time. A fundamental role is played by the influx of resource
into the system (e.g. glucose) and how the resource is handled by the population, which in its turn can increase or shrink
in size according to the amount of resource provided. In every time step, both individuals and groups can give
rise to new individuals and groups, respectively. Each individual
divides into two identical cells as soon as its internal energy storage $E_{ij}$ (group index, $i=1,..., N_G$, and individual index, $j=1,...,P_i$) surpasses 
the energy threshold $E_{\rm{max}}$. Each daughter cell inherits half of the energy of the parental cell.
Groups have a threshold size $P_{\rm{max}}$, and whenever the number of individuals in a group reaches this limit, the group
splits into two groups, with its members being evenly distributed between the two.
The above described dynamics is also illustrated in Fig.\,1.
In addition, each individual dies at a constant probability $\nu$ per time step. The group death occurs 
when the number of individuals shrinks to zero.

\bigskip
In every time step, the amount of resource available to the population is constant and equal to $S$. We consider equipartition of 
resource among the groups and so the amount of resource per group is $S_G=S/N_G$. Note that the total number of groups $N_G$ is a 
dynamical number.
We break the complete process of an individual gaining energy in two stages: first, individuals compete for the resource;
second, the caught resource is ultimately converted into internal energy (ATP). The assumption of breaking down 
the process in two stages is in accordance with mechanistic models of cells \shortcite{WeibePNAS2015}: first, an enzyme 
transports  the resource into the cell, and then a second set of enzymes metabolizes the 
resource into a generic form of energy, which includes ATP. The 
rates of import and of metabolism are distinct \shortcite{WeibePNAS2015}. 
\\
\\
\noindent
{\bf Implementing the resource uptake}
\\
\\
\noindent
The resource is shared among the members of a group. As a strain type with rapid metabolism, $(D)$, has a higher 
rate of consumption, it seizes a larger portion of the resource available to the group compared with 
strains type $C$, characterized by a low consumption rate. The amount of resource per strain $D$ in a given group $i$ 
is estimated as 
\begin{align}\label{J}
S^i_D(S_G) = \frac{J^S_D(S_G)}{J^S_D(S_G)P^i_D+J^S_C(S_G) P^i_{C}}\ S_G,
\end{align}
where $P^i_D$ and $P^i_C$ are, respectively, the number of individuals for type $D$ and $C$ in group $i$, such 
that $P^i_D + P^i_C = P^i$ corresponds to the size of group $i$. 
The consumption rates for strain type $C$ and $D$, $J^S_C(S_G)$ and $J^S_D(S_G)$, are 
functions of $S_G$, the amount of resource per group.
Similarly, the amount of resource per strain C is 
\begin{align}\label{JATP}
S^i_C(S_G) = \frac{J^S_C(S_G)}{J^S_D(S_G)P^i_D+J^S_C(S_G) P^i_{C}}\ S_G.
\end{align}
Therefore, the total amount of resource for group $i$ is $S^i_D(S_G)P^i_ D + S^i_C(S_G)P^i_C=S_G$ as declared above.
\\
\\
\noindent
{\bf Implementing the conversion of resource into internal energy}
\\
\\
\noindent
The resource inside the cell is used to increment its internal energy $E_{ij}$, which is calculated as
\begin{align}\label{energy}
\Delta E_{ij} = J^{ATP}_C(S^i_C) \text{ or } \Delta E_{ij} = J^{ATP}_D(S^i_D),
\end{align}
depending on strain type, $C$ or $D$. The functions $J_C^{ATP}(S^i_C)$ and $J_D^{ATP}(S^i_D)$ represent 
how efficiently the resource ($S^i_C$ or $S^i_D$) is transformed into energy (ATP) for each individual type.

\subsection{Simulation procedure}

\subsubsection{Statistical measurement}
In the most probable scenario for the establishment of the efficient mode of metabolism, an alternative metabolic pathway
arose by mutation in a pool of organisms making usage of resource inefficiently. For this sake, it is of practical
importance to determine the probability that a single individual of type $C$ invades 
and ultimately dominates a population.
The simulation initiates from an isogenic population of strains $D$ and let it evolve 
to a stationary regime. This step is required as the population dynamics and the resource availability determine the population size at the steady regime. For this reason the system
is said to be self-organized.  At this point, a single strain $C$ replaces a randomly chosen cell from the
population. From this time on, the fate of the mutant type $C$ is observed, i.e., the simulation is tracked until
the mutant gets lost or fixed. 
\bigskip

The fixation probability is simply the fraction of independent runs at which the efficient strain gets fixed. 
Because the population size itself is an outcome of the dynamics of the model, it is rather desirable to present
the outcomes in terms of the relative fixation probability, defined as fixation probability 
divided by $1/N_{\rm{st}}$. Here $N_{\rm{st}}$ stands for the population size of strains $D$ at the moment the efficient strain is introduced. Thus, $1/N_{\rm{st}}$ corresponds to
the fixation probability of a mutant under neutral selection in a population of size $N_{\rm{st}}$.  
If the relative fixation probability is larger than one, the mutant 
type is said to be selected for, while a probability smaller than one means the strain is counter-selected.

\subsubsection{Parametrization of the model}
The rates of resource uptake and of conversion of food into energy appearing in Eqs. (\ref{J}-\ref{JATP}) are defined as
\begin{subequations}\label{parametrization} 
\begin{align}
	&J^S_C(S_G) = A_C\left(1-\exp(-\alpha{S_G})\right) \\
	&J^S_D (S_G)= A_D\left(1-\exp(-\alpha{S_G})\right)\\
	&J^{ATP}_C(S_i^C) = A^{ATP}_C\left(1-\exp(-\alpha^{ATP}_C{S_C^i})\right)\\
	&J^{ATP}_D (S_i^D)= A^{ATP}_D\left(1-\exp(-\alpha^{ATP}_D{S_D^i})\right).
\end{align}
\end{subequations}
In common, all these functions display a sigmoidal shape, thus saturating for large values of $S$, similarly
to Michaelis-Menten functions \cite{PfeifferScience2001,WeibePNAS2015}. The effect of the rates on the population dynamics
is here extensively studied. For most of the parameters we will perform a sweep over the parameter space that covers
the physically meaningful regions of interest.  
\noindent
The maximum values for the rates are chosen in such way that the rate-yield trade-off exists. 
By definition, the efficient strain $C$ is the one that transforms resources into internal energy efficiently 
at the expense of the consumption rate, while strain $D$ metabolizes fast but at low yield. The parameters above follow
some conditions: 
\begin{itemize}
\item[(i)] $A_D$ must be larger than $A_C$  ($A_D > A_C$), which assures that type $D$ always 
has a higher consumption rate for the same amount of resource.
Note that the argument in the exponential function is exactly the same as in 
Eqs. \ref{parametrization}a and \ref{parametrization}b, and so it will have no influence on the dynamics. We just keep it
for future generalisations. As naturally emerging in the forthcoming calculations, the relevant parameter is in fact the
ratio between these two quantities, $\epsilon=A_{D}/A_{C}$.
\item[(ii)] Another condition is $\Delta_{ATP}=\alpha^{ATP}_D/\alpha^{ATP}_C < 1$. Once again, the ratio between the exponents in the
functions (\ref{parametrization})c and  (\ref{parametrization})d arises as a key quantity in the process. 
This condition warrants that strain $D$ is less effective
than strain $C$ by generating energy from a given amount of resource, unless a very large amount of resource is gathered to be converted
into energy. This is ensured when the ratio $\Gamma_{ATP} = A^{ATP}_D/A^{ATP}_C$ is larger than one, which allows a fast metabolic pathway
to produce more energy in case they capture a much higher amount of resource than the efficient strain. 
This possibility is particularly important since it is known that many cells can work as a respiro-fermenting cells, concomitantly 
using the two alternative pathways of ATP production. Such respiro-fermentative metabolism is a typical mode 
of ATP production in unicellular eukaryotes like yeast 
 \shortcite{FrickNatur2003,OtterstedtEMBO2004}. In such a situation, the yield of the whole process is reduced
as the cell drives more resource to be metabolised inefficiently, a habitual behavior under the condition of plentiful
resource. Of course, the situation $A^{ATP}_D < A^{ATP}_C$ can also be seen, but the 
opposite $A^{ATP}_D > A^{ATP}_C$ represents the worst scenario for the efficient 
strain, and if it can thrive in such case it will naturally be favored under less harsh
conditions.
\end{itemize}

\subsection{Social conflict}
Besides settling down the constraints on the biological relevant parameter values, it is also important to find the parameter region at which the rate-yield trade-off in fact produces a tragedy of the commons \cite{HardinScience1968}. 
The social conflict exists if in a pairwise competition for a common resource the strain with fast metabolism 
outcompetes the strain with high yield \shortcite{GudeljJEVOLBIOL2007,PfeifferScience2001,MacleanHeredity2008}.

\bigskip
For a given resource influx rate $S^{*}$, the uptake of resource by strains $C$ and $D$ are, respectively, 
\begin{align}
&S_{C}  =  \frac{J_{C}^{S}}{J_{C}^{S}+J_{D}^{S}}S^{*} \\ 
&S_{D}  =  \frac{J_{D}^{S}}{J_{C}^{S}+J_{D}^{S}}S^{*}.
\end{align}  
The ratio between $J_{D}^{ATP}$ and $J_{C}^{ATP}$ provides us the relative advantage of  strain $D$ over
strain $C$. 
From Eqs. (\ref{parametrization}) we obtain that
\begin{equation}\label{social}
r=\frac{J_{D}^{ATP}}{J_{C}^{ATP}}=\Gamma_{ATP}\frac{1-\exp(-\alpha_{D}^{ATP}S_{D})}{1-\exp(-\alpha_{C}^{ATP}S_{C})}.
\end{equation}
If this ratio is larger than one, the social conflict is held true. This reasoning will be employed to 
map the set of parameter values at which social conflict exists. As the ratio $r$ depends on the amount of resource, in order
to delimit the social conflict region in a population of arbitrary size,  we must use $S^{*} \simeq S/N$, instead of using $S$, the
 total resource influx rate. $N$ denotes the population size. Though, as observed from the simulations, 
the system evolves in such way that $S/N$ is always small at equilibrium.  As we will see, the limit of small $S/N$ will prove 
to be useful in the forthcoming.
\section{Results}
\label{sec:results}
We will start this section by presenting some analytical approximations for well-mixed populations. 
These calculations are useful in the sequel and serve as a guidance to determine 
the situations at which the simulations results match the theoretical prediction. 
They also help us to delimit the situations where phenomena driven by stochastic effects manifest, 
which are not captured by the analytical approach carried out here. 
As in the analytical approach structuring is not assumed, it will be particularly useful to
demonstrate whether group structure can be a driving mechanism for the maintenance of the efficient strain.

\subsection{Analytical results for well-mixed populations}
\subsubsection{Equilibrium size of a population of strains with fast metabolism}
\label{sec:equilibD}
Let us propose a discrete-time model for a well-mixed population with only cells of type $D$.
More generally, the population size in time step $t+1$ can be written as
\begin{equation}
n(t+1)=n(t)+ g(n(t),S)\,n(t) - \nu\, n(t),
\label{EqPopSize}
\end{equation}
where the function $g(n(t),S)$ denotes the growth rate of strain $D$, which depends on the population size as well as on 
the resource influx rate, $S$.  The death rate is denoted by $\nu$ (as a mathematical guide for the analytical development to be done, we
suggest as a reference \shortcite{DayBook}).  
In the above equation, it is implicitly assumed that $g(n(t),S)$ is proportional to the amount of resource captured by the consumer. 
The conversion of resource into energy (directly translated as growth rate) occurs into 
two steps: uptake of resource, and its posterior conversion into energy. 
As there is only one type of strain, the amount of resource per individual at time $t$ 
is just $S/n(t)$.  Following Eq.\,(\ref{parametrization}d), the growth rate is equivalent to
$g(n(t),S)=a_{D}\left[1-\exp(-\alpha_{D}^{ATP}S/n(t))\right]$, where $a_{D}=A_{D}^{ATP}/E_{\rm{max}}$, which is a necessary rescaling 
because individuals only replicate if their internal energies surpass the threshold $E_{\rm{max}}$. Plugging this expression into 
Eq. (\ref{EqPopSize}) we get 
\begin{equation}\label{eq1}
n(t+1)=n(t)+a_{D}\left[1-\exp(-\alpha_{D}^{ATP}S/n(t))\right] n(t)-\nu n(t).
\end{equation}
The system is under equilibrium when $n(t+1)=n(t)=\hat{n}$ and has two 
equilibrium solutions, $\hat{n}=0$ and $\hat{n}=-\frac{\alpha_{D}^{ATP}S}{\ln (1-\nu /a_{D})}$. Note 
that the second solution is only meaningful if $\nu < a_{D}$.

\subsubsection{Stability Analysis}
Next we analyze the local stability of the two equilibrium solutions. By writing the dynamics of the discrete-time model 
like $n(t+1)=f(n(t))$, a solution, $n=\hat{n}$, is said to be stable when $-1 <\lambda < 1$, where 
$\left. \lambda = \frac{df}{dn}\right|_{n=\hat{n}}$. Note that the derivative of $f(n(t))$ with respect to $n$ must 
be evaluated at the equilibrium. A value of $\left| \lambda \right|<1$ warrants that any small perturbation from the equilibrium 
solution will be damped, bringing the system back to equilibrium. Whenever $\left| \lambda \right|>1$, any disturbance 
drives the system away from the equilibrium point.

\bigskip
For the first equilibrium, $\hat{n}=0$, $\lambda$ is evaluated as $\left. \lambda=\frac{df}{dn}\right|_{\hat{n}=0}=1-\nu+a_{D}$, and so
the solution is stable when $\nu>a_{D}$, i.e. $\nu E_{\rm{max}}>A_D^{ATP}$.  On the other hand, the second 
equilibrium, $\hat{n}=-\frac{\alpha^{ATP}S}{\ln (1-\nu/a_{D})}$, provides $\left. \lambda = \frac{df}{dn}\right|_{\hat{n}=-\frac{\alpha^{ATP}_D\,S}{\ln (1-\nu/a_{D})}}=1+a_{D}(1-\nu/a_{D})\ln (1-\nu/a_{D})$. Because $\nu<a_{D}$ in the range of validity of the solution, the logarithm 
is always negative, meaning $\lambda < 1$. Therefore, the second solution is
always stable in the range it is valid.

\subsection{Evolutionary Invasion Analysis}
\subsubsection{Invasion by the efficient strain} 
\label{section:invasion by the efficient strain}
A further step of our analytical calculation is to check under which conditions a population of strain with rapid growth 
can be invaded by a mutant that uses the resource more efficiently. 
The simulations depart from an isogenic population of strain $D$, and after a stationary state a single type $C$ individual is introduced. The dynamics in terms of the number of individuals of strains $D$ and $C$ over time, $n_{D}$ and $n_{C}$ respectively, 
is now determined  by the following set of equations:
\begin{align}\label{eq-defcop}
n_{D}(t+1)&=n_{D}(t)\{1+a_{D}\left[1-e^{-\alpha_{D}^{ATP}\frac{J_{D}^{S}S}{J_{D}^{S}n_{D}+J_{C}^{S}n_{C}}}\right]-\nu\} \\
n_{C}(t+1)&=n_{C}(t)\{1+a_{C}\left[1-e^{-\alpha_{C}^{ATP}\frac{J_{C}^{S}S}{J_{D}^{S}n_{D}+J_{C}^{S}n_{C}}}\right]-\nu\}, 
\end{align}
and now $a_{C}=A_{C}^{ATP}/E_{\rm{max}}$. It is important to point out that since there are two distinct strains, the
amount of resource captured per individual is no longer  $S/n$ but obeys the resource partition as calculated
by Eqs.\,\ref{J}$\&$\ref{JATP}. The dynamics is driven by a density dependent growth rate. 
One possible equilibrium of the above system is
\begin{equation}\label{solution-nd}
\hat{n}_{D}=-\frac{\alpha_{D}^{ATP}S}{\ln (1-\nu/a_{D})}   ~ ~~~~~~~ \mbox{and} ~~~~~~~~ \hat{n}_{C}=0.
\end{equation}
This equilibrium corresponds to the nonzero equilibrium of the resident population of strain $D$, as shown 
in \ref{sec:equilibD}. 

\bigskip
Next we analyse the stability conditions of this equilibrium under the invasion of a small quantity of efficient strains. 
For multiple-variable models, the local stability of the equilibrium solutions is found
by looking at the eigenvalues of the Jacobian matrix of the system, see details in the Supporting Information. 

\bigskip
After determining the eigenvalues and setting the conditions for stability (please see Supporting Information),  it
is possible to disclose a relationship delimiting the parameter region at which the efficient strain can invade the resident population of type $D$ cells. For fixed $A_{C}^{ATP}$ 
and variable $A_{D}^{ATP}$ we find that
\begin{equation}\label{stability}
\Gamma_{ATP}=\frac{\nu/a_{C}}{1-(1-\nu/a_{C})^{\Delta_{ATP}\epsilon}},
\end{equation}
where $\Delta_{ATP}=\frac{\alpha_{D}^{ATP}}{\alpha_{C}^{ATP}}$ and $\epsilon=\frac{A_{D}}{A_{C}}$, as previously defined. 
Eq.(\ref{stability}) is used for the comparison with the isocline $P_{fix}=1$, fixation probability equal to one, obtained from 
simulations. Therefore, our finding shows that the solution, eq.(\ref{solution-nd}),
is stable against invasion above the curve (\ref{stability}), while the strain $C$ can succeed to invade
in the parameter space under the curve.

\subsubsection{Invasion by the inefficient strain}
\label{section:invasion by the inefficient strain}
Now the opposite situation is considered --- the conditions 
under which a mutant with rapid metabolism (type $D$) can invade
a population dominated by cells of type $C$. The calculations are straightforward as calculations are similar as in Sec.\ref{section:invasion by the efficient strain}.
As the resident population is initially composed of cells type $C$ only, its size at equilibrium 
is $n_{c}=\hat{n}_{C}=-\frac{\alpha_{C}^{ATP}S}{\ln (1-\nu/a_{C})}$, and $a_{C}=A_{C}^{ATP}/E_{\rm{max}}$.

\bigskip
The Jacobian matrix is now evaluated at 
\begin{equation}\label{solution_nc}
\hat{n}_{C}=-\frac{\alpha_{C}^{ATP}S}{\ln (1-\nu/a_{C})} ~ ~~~~~~~ \mbox{and} ~~~~~~~~ \hat{n}_{D}=0
\end{equation}
as the strain $D$ is assumed to be rare. Likewise, a relation between $\Gamma_{ATP}$ and $\Delta_{ATP}$ is obtained, and
the curve now delimits the parameter region at which the efficient strain 
is stable against the invasion by the inefficient strain. It is found that 
\begin{equation}\label{stability_cop}
\Gamma_{ATP}=\frac{\nu/a_{C}}{1-(1-\nu/a_{C})^{\Delta_{ATP}\epsilon}},
\end{equation}
which is exactly equal to Eq.(\ref{stability}). However, now the solution, eq. (\ref{solution_nc}), is stable under the curve. 
Thus, we conclude that the parameter region $\Gamma_{ATP}$ versus $\Delta_{ATP}$, where the efficient 
strain is evolutionary stable against the invasion by the selfish strain $D$, coincides with 
the region at which it can, while rare, invade an isogenic population of type $D$ individuals. 

\subsection{Simulation Results}
In this section we present the simulation results of the evolutionary invasion analysis. 
In Supporting Information we provide simulation results and present the plots of the population size
of isogenic population of strain  $D$ at equilibrium. This is critical as the population size determines the strength of stochasticity on the dynamics of the system.

\begin{figure}[htp]
\centering
\includegraphics[width=\textwidth]{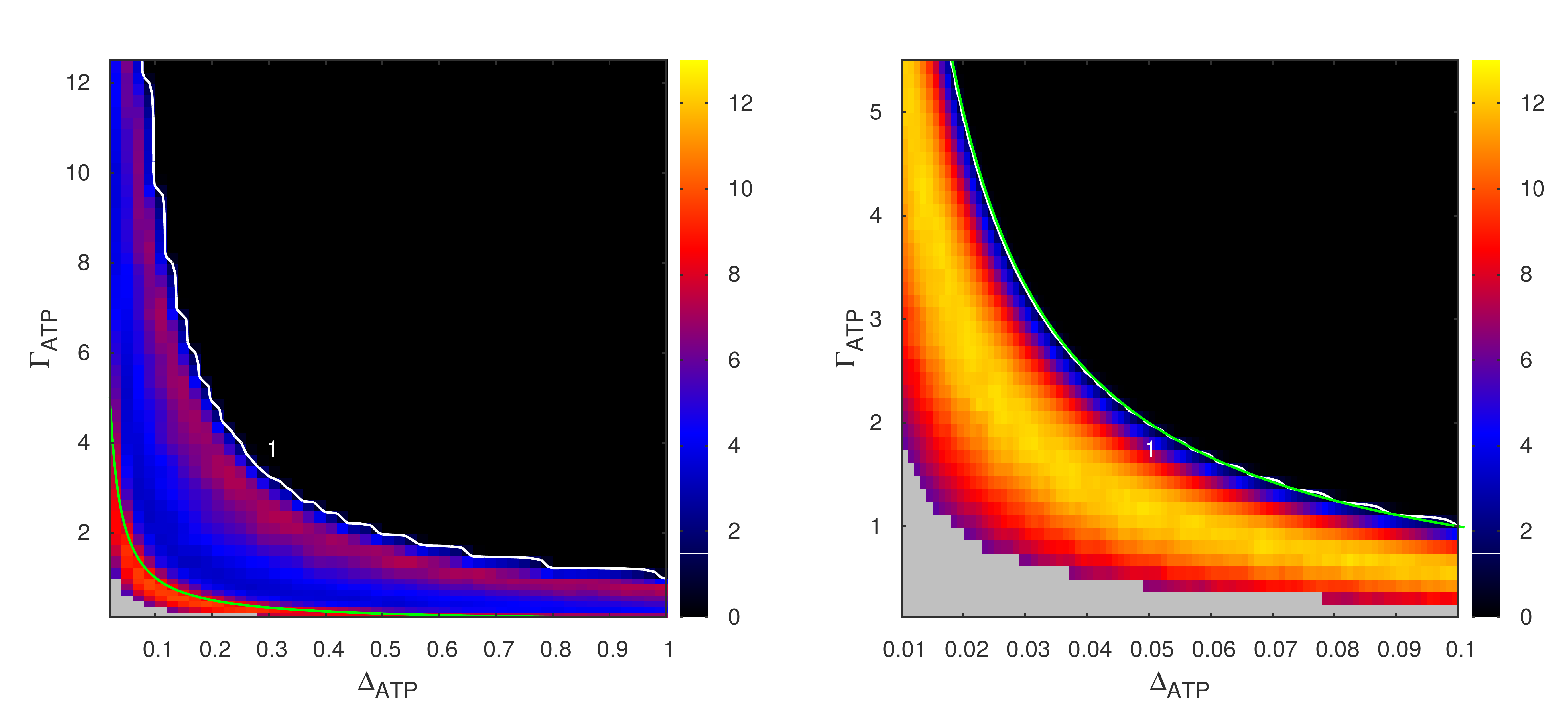}
\caption{Fixation probability. In the plot the fixation probability is shown in terms of $\Gamma_{ATP}=A^{ATP}_D/A^{ATP}_C$
 and $\Delta_{ATP}=\alpha^{ATP}_D/\alpha^{ATP}_C$. Left panel: structured populations, right panel: homogeneous populations. The white thick 
lines denote the isocline where the fixation probability of a single cooperator is the same as the fixation probability of a random individual under neutral selection. Above the isocline the cooperative strategy is counter-selected (dark region), whereas under the line
it is selected for. Yet, the green line corresponds to the line delimiting the social conflict regime, obtained by making $r=1$ in eq. \ref{social}. This line overlaps the lines given in eqs (\ref{stability}) and (\ref{stability_cop}). Once again, the grey 
region denotes that the defector population goes to extinction before the cooperator is introduced.
The other parameter values are influx rate of resource $S=25$, death 
rate $\nu=0.01$, group carrying capacity $P_{\rm{max}}=10$, internal energy for
splitting $E_{\rm{max}}=10$, and $A_{D}=10$.  The 
data points corresponds to $40$ distinct populations and for each population $10,000$ independent runs were performed.}
\end{figure}

\subsubsection{\small{The relative fixation probability of a single individual of the efficient strain}}
From the heat map in Fig.\,2 we observe how the fixation probability is influenced by the 
ratios, $\Gamma_{ATP}=A^{ATP}_D/A^{ATP}_C$ and $\Delta_{ATP}=\alpha^{ATP}_D/\alpha^{ATP}_C$. The ratio $\epsilon =A_{D}/A_{C}$ is set at 
$10$, which is a typical value, as empirically measured in populations of {\it S. cerevisiae} 
\shortcite{OtterstedtEMBO2004}. Making the comparison with well-mixed populations, 
over a broader domain of the parameter space the efficient strain is selected for 
in the structured populations. When $\Delta_{ATP}$ is already close to one, meaning that 
the yield of strain $D$ is almost equivalent to strain $C$, it still persists 
a domain at which efficient strain is favored. The domain 
at which the efficient strain is favored in well-mixed populations is considerably restricted 
in comparison to structured populations. For instance, around $\Delta_{ATP}=0.1$
the efficient strain is already unlikely to invade and to fixate. 

\bigskip
The curves corresponding to eqs. (\ref{stability}) and (\ref{stability_cop}) are represented by the green line in the left and the 
right panels of Fig.\,2.  The region below the green line portrays the domain at which
the efficient strain can invade and replace a population of individuals type $D$. Of course, the outcome of the
dynamics is not deterministic, but it evidences a selective advantage for strain  $C$ .  
The white line is the isocline $P_{fix}=1$ (fixation probability is equal to one). The line setting out the onset 
of the social conflict, corresponding to $r=1$ in eq. (\ref{social}), is also plotted. In well-mixed populations (right panel of Fig.\,2) 
the social conflict line ($r=1$) overlaps with the green line, eqs. (\ref{stability}) and (\ref{stability_cop}), and thence 
it is not perceivable in the plot. Indeed, in the limit $S/N$ small, which always holds when $\nu/a_{D}$ is also 
small (see Eq. (\ref{solution-nd})), the two curves are essentially the same and become $\Gamma_{ATP}=\frac{1}{\Delta_{ATP}\epsilon}$. 
This observation leads us to conclude 
that in well-mixed populations the efficient strain is an effective intruder only 
in the domain there is no dilemma. 

\bigskip
On the other hand, in structured populations the efficient strain is selectively 
advantageous over a much larger set of the parameter space. The realm of the high yield strain
goes further beyond the line setting out the social conflict. The isocline 
$P_{fix}=1$ is considerably shifted to the right. The region  
between the social conflict line (green line) and the isocline $P_{fix}=1$ is of special interest, since it shows
the prominent role of structuring --- more specifically kin-selection --- on promoting the fixation
and maintenance of the strain exploiting the resource efficiently. 

\begin{figure}[htp]
\centering
\includegraphics[width=\textwidth]{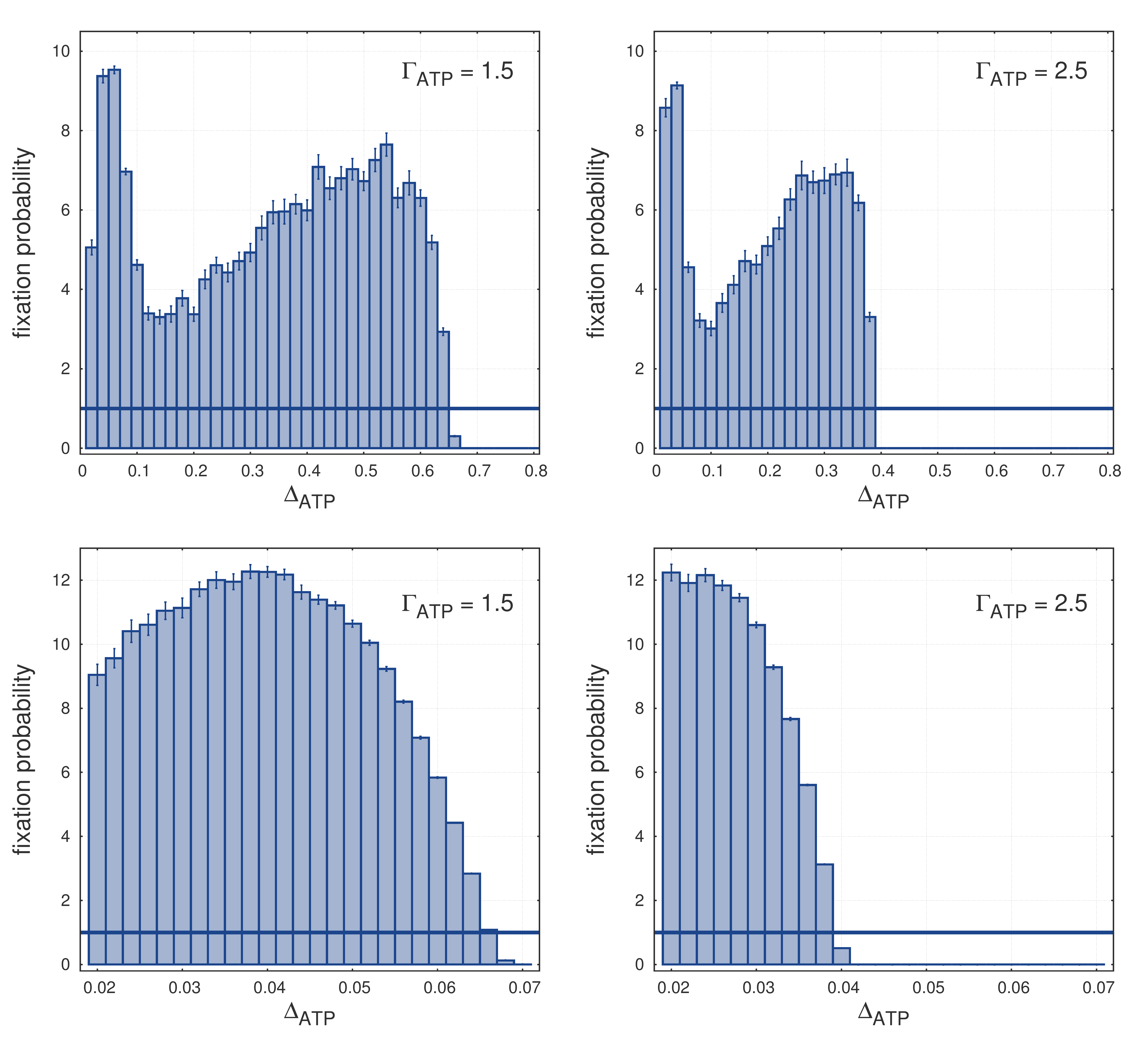}
\caption{Fixation probability as a function of $\Delta_{ATP}$ for fixed values of $\Gamma_{ATP}$. 
In top panel we exhibit the simulation
outcomes for structured populations whereas in the bottom panels for homogeneous populations. 
The data points correspond to $40$ distinct populations and for each population $100,000$ independent runs were performed.
The parameter values are $S=25$, $\nu=0.01$, $P_{\rm{max}}=10$, $E_{\rm{max}}=10$, $A_{D}=10$, and $A_{C}=1$.}
\end{figure}

\bigskip
Looking at the heat map, it is noticeable that there exists a non-straight\-for\-ward relationship between
$\Delta_{ATP}$ and $\Gamma_{ATP}$. At fixed $\Gamma_{ATP}$ the relative fixation probability approaches a two-humped function of
$\Delta_{ATP}$, being maximised at low and intermediate values of $\Delta_{ATP}$ (see Fig.\,3). In the panels, $\Gamma_{ATP}$ is kept constant
while $\Delta_{ATP}$ varies. Well-mixed populations (lower panels) display a smoother scenario and the relative fixation
probability is now an one-humped function of $\Delta_{ATP}$. The point, at which the relative probability is maximised, depends on 
$\Gamma_{ATP}$, being shifted towards higher values of the ratio $\Delta_{ATP}$ as $\Gamma_{ATP}$ decreases. 
In well-mixed populations the selective advantage of strain $C$ over strain $D$ is confined to the
region where there is no dilemma. 
Yet, in structured populations the range of $\Delta_{ATP}$ in the plot embodies three distinct regions: the region at which there
is no dilemma, a second region where the dilemma exists and the efficient strain has a selective advantage, and a last one at which
the dilemma still holds but the efficient strain is no longer able to persist. The first peak occurs at small  $\Delta_{ATP}$ as
observed in well-mixed population, but now there is a second peak that lies at the range
of intermediate values of $\Delta_{ATP}$. This latter peak has no counterpart in a 
scenario of well-mixing, as it is promoted by  kin-selection.

\begin{figure}[tp]
\centering
\includegraphics[width=\textwidth]{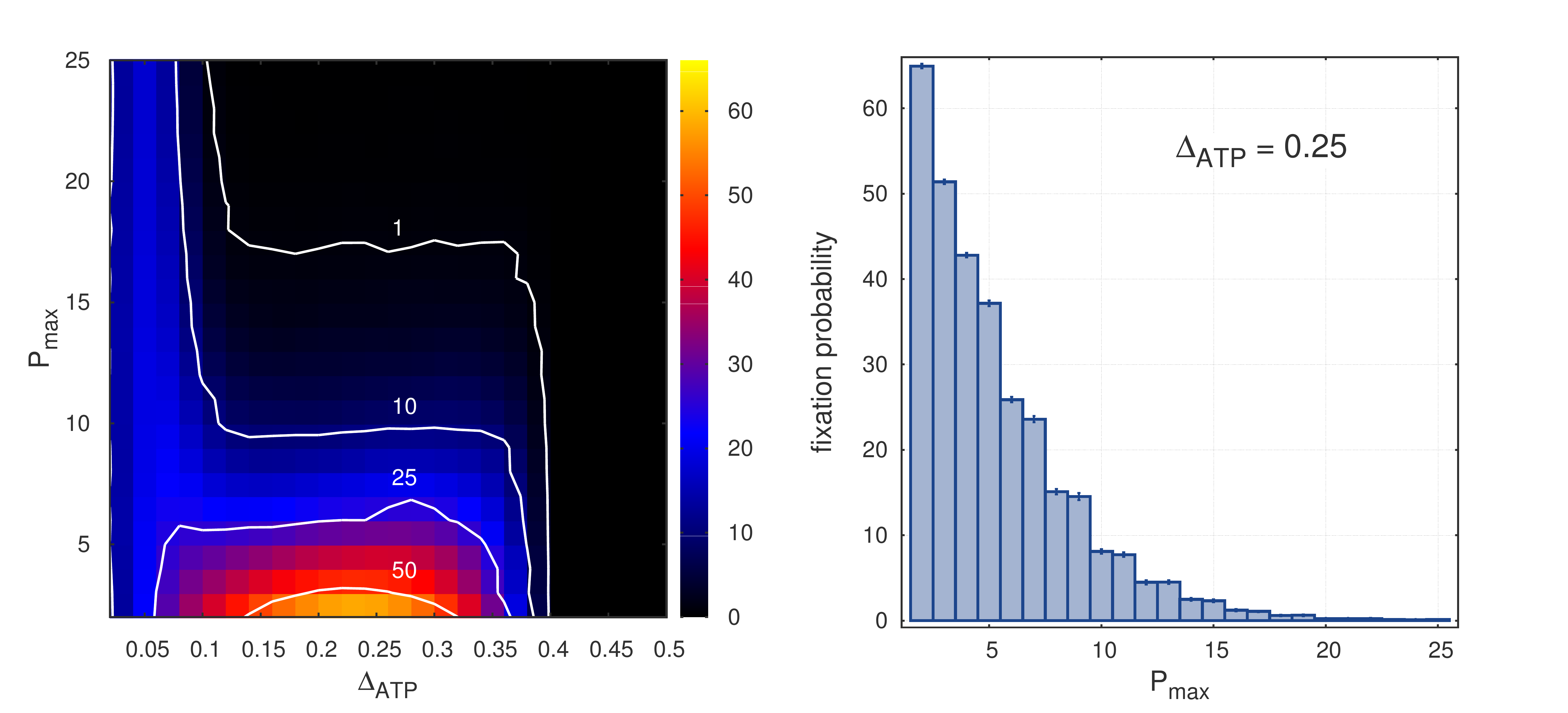}
\caption{The strength of stochasticity. The left panel shows the fixation probability in terms of the carrying capacity $P_{\rm{max}}$ and
the ratio $\Delta_{ATP}$. The right panel represents how the fixation probability changes with $P_{\rm{max}}$ in a given $\Delta_{ATP}$. 
The thick white lines correspond to isoclines. The parameter values are $S=25$, $\nu=0.01$, $E_{\rm{max}}=10$, $A_{D}=5$, and $\Gamma_{ATP}=2.5$. The data points correspond to $40$ distinct populations and for each population $10,000$ ($100,000$) independent runs were performed
in left panel (right panel).}
\end{figure}

\bigskip
Next we study how the relative fixation probability changes with the group carrying capacity, $P_{\rm{max}}$.
This analysis enables to assess the strength of stochasticity and kin-selection on the population dynamics. 
Smaller group sizes entail higher relatedness among
individuals in the same group. Previous studies have claimed that in multilevel selection the likelihood of 
invasion and posterior fixation of cooperative traits is particularly enhanced for small group sizes \shortcite{NowakPNAS2006}, 
evidencing the importance of stochasticity in the dynamics \shortcite{NowakNature2004}. 
Similar behaviour was observed in a resource-based model of pairwise interactions within the 
multilevel selection framework \shortcite{FerCampos2013}.
The contour map in the left panel of Fig.\,4  displays the
relative fixation probability for 
different values of $P_{\rm{max}}$ and different values of $\Delta_{ATP}$. 
The lighter colours in the bottom of the plot evince that 
the efficient strain $C$ is significantly favored as smaller group size is considered. 
For extremely small group sizes, like $P_{\rm{max}}=5$, the fixation probability varies enormously with the ratio
$\Delta_{ATP}$, being maximised at intermediate values. In such small groups, the fixation probability can be  
sixty-fold higher than fixation probability of a neutral trait. At large values of $P_{\rm{max}}$, instead of a pronounced change
with $\Delta_{ATP}$, the probability is roughly constant at a broad interval range of $\Delta_{ATP}$, as 
inferred from the isoclines.
Of particular interest is the isocline delimiting the onset of the phase at which 
the efficient strain is favored. From the isocline $P_{fix}=1$ one deduces 
that for $\Delta_{ATP}$ varying between $0.1$ and $0.4$
there is a well-defined critical value for the group carrying capacity, which is about $P_{\rm{max,c}} \simeq  17$,
beyond which the efficient strain becomes counter-selected.  
Though, as the efficiency of  the strain $D$ 
is further reduced (small $\Delta_{ATP}$), the critical group size $P_{\rm{max,c}}$ soars again. 
For the sake of completeness, the right panel of Fig.\,4 shows that the 
relative fixation probability is a well-behaved monotonic 
decreasing function of $P_{\rm{max}}$ under fixed $\Delta_{ATP}$. 

\begin{figure}[tp]
\centering
\includegraphics[width=0.5\textwidth]{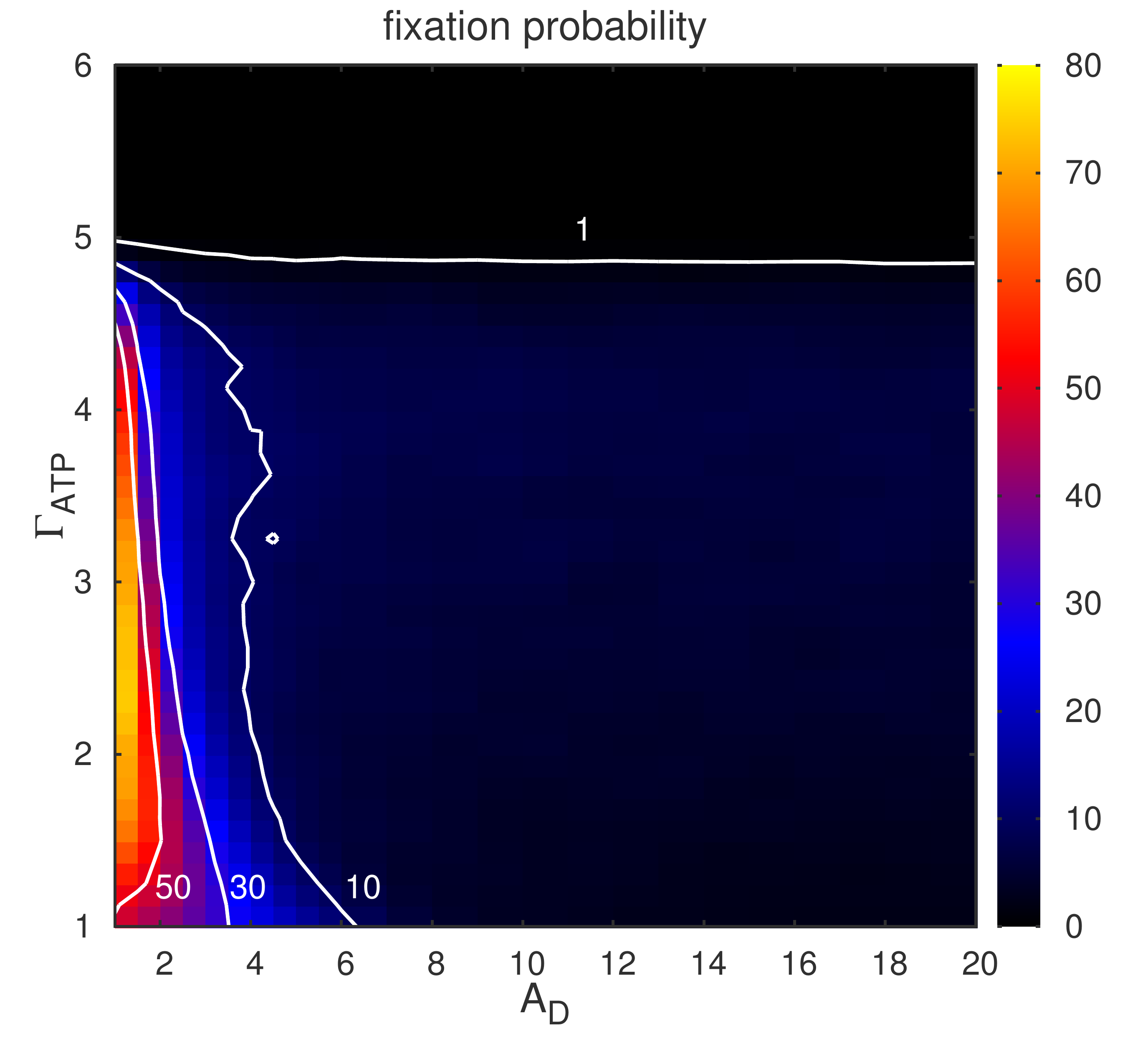}
\caption{The effect of the consumption rate. Heat map of the fixation 
probability in terms of the ratio $\Gamma_{ATP}$  and
the consumption rate of defectors $A_{D}$. The parameter values are $S=25$, 
$\nu=0.01$, $P_{\rm{max}}=10$, $E_{\rm{max}}=10$, and $\Delta_{ATP}=0.2$. 
The thick white lines correspond to isoclines. The data points correspond 
to an average over $10$ distinct populations and for each population $10,000
$ independent runs.}
\end{figure}

\bigskip
How the consumption rate affects the fate of the efficient strain with a low frequency is studied in Fig.\,5. A contour plot
in terms of $\Gamma_{ATP}$ and $A_{D}$ is presented. 
The consumption rate is naturally a relevant part in the process since the amount 
of energy added to the individual's internal energy is bounded to the amount of resource caught, which 
by its turn also influences other individuals' performance due to competition.
The selfish strategy enjoys a twofold benefit when both $A_{D}$ and $\Gamma_{ATP}$ are kept high.  First in the competition for
the resource, as a high $A_{D}$ means a big advantage in the resource uptake, and subsequently in process of conversion of resource into energy
which is determined by its efficiency, see Fig.\,5. As 
illustrated in the plot, the left right corner is
dark-coloured meaning that the efficient strain is clearly counter-selected. 
$\Gamma_{ATP}$ has a more influential role
than $A_{D}$. At $A_{D} \simeq 5$, a further increase of the variable does not lead to any substantial variation of the
fixation probability.

\subsection{Enabling migration between groups}
As learned from the previous discussions, a way to reduce the strength of kin-selection and thus allowing 
a less favorable environment  for the promotion of efficient usage of resource is to increase 
the group carrying capacity $P_{max}$. An alternative manner to change the strength of kin-selection is presented.
Now we allow individuals to migrate: at each generation every individual has a probability $m$ to leave its original group
and move to a randomly chosen group in the population. This is known as  the island model of stochastic 
migration \cite{WrightGenetics1943}. By allowing migration to occur, one can tune the level of relatedness among individuals, which tends
to decrease as the migration rate is increased. 

\begin{figure}[tp]
\centering
\includegraphics[width=\textwidth]{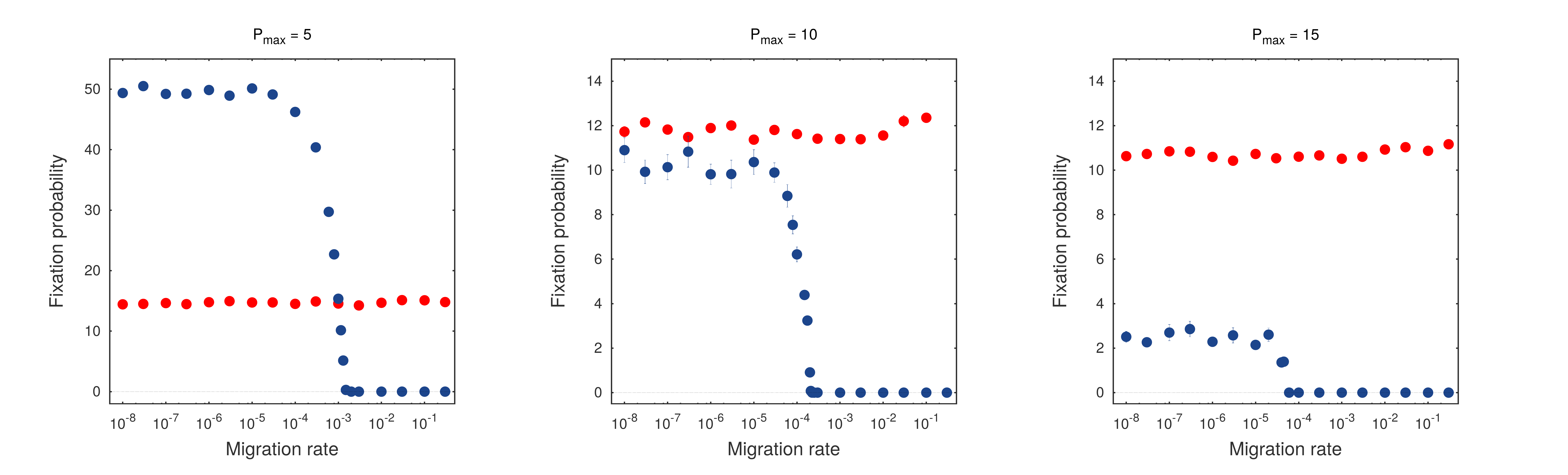}
\caption{Effect of migration on the fixation probability. The fixation probability is plotted as a 
function of the migration rate $m$ for three distinct values of carrying capacity $P_{\rm{max}}$. 
The parameter values are $S=25$, $\nu=0.01$, $E_{\rm{max}}=10$, $A_{D}=10$, $A_{C}=1$, $\Gamma_{ATP}=1.5$ and $\Delta_{ATP}=0.5$ (blue points) 
and $\Delta_{ATP}=0.04$ (red points).}
\end{figure}

\bigskip
The effect of the migration rate on the fixation probability can be seen in Fig.\,6. 
Different situations are considered in the face of our previous knowledge about the system. 
Distinct values of $\Delta_{ATP}$ and of carrying capacity $P_{\rm{max}}$ are considered.
At $\Delta_{ATP} = 0.5$ (blue points) the efficient strain is selected against
in well-mixed populations but not in structured populations.
At $\Delta_{ATP} = 0.04$  (red points) the efficient strain is favored in both scenarios.  
In the latter situation, migration does not affect the outcome and the fixation probability is invariant.
Helpful information about the role of migration is gathered from the analysis of the case $\Delta_{ATP} = 0.5$.
Starting from small migration rates, which means the groups are more isolated and relatedness is high, the 
fixation probability is nearly the same over a wide range of
migration, and then sharply goes to zero as the rate of migration is further augmented, signifying less isolation. The 
first plateau coincides
with the value of the  fixation probability in our structured population model, while after 
the transition the situation resembles well-mixed populations at which the efficient strain can no longer thrive. 

\section{Discussion}
\label{sec:conclusions}
We have studied the evolutionary dynamics of a population with two strains under a trade-off between rate and yield.
The two strains with distinct metabolic pathways compete under a scenario of
limited resource. This problem has long been debated within the framework of evolution of cooperation, as
the strain that uses resource efficiently is described as a cooperator, while the strain with high uptake rate,  at the 
expense and degradation of its own environment, is said to display a defecting
behavior \shortcite{PfeifferScience2001,kreftMicrobiology2005}. The trade-off between high rate and yield occurs due to
biophysical restraints. Experimental results show a clear negative correlation between resource uptake and
efficiency of conversion of resource into energy \shortcite{PfeifferScience2001,KreftMicro2004,MeyerNatComm2015}. The existence 
of such trade-offs
in heterotrophic microorganisms, like bacteria and yeast,  and the fact 
that the efficient manipulation of resource was made possible with the appearance of the
metabolic pathway that makes use of oxygen, the so-called respirers,  have motivated a large body of studies.
A clear motivation is the issue related to the major transition 
from single cells towards multicellular forms of life \shortcite{PfeifferPNAS2003,AledoJMolEvol2008}. Multicellularity can 
thus be thought as the integration of
smaller units to build up a more complex machinery.

\bigskip
Kin selection as one possible driving force promoting the units that make efficient handling of resource 
is effective when relatedness among individuals is relatively high such that the cost of the cooperative act can be balanced. 
A natural way to enhance relatedness is to increase cohesion or the level of clustering in the population. Here we proposed
an explicit resource-based model in which competing metabolic strategies are assumed. We studied the conditions under which
the cooperative strain with high yield metabolism can invade and outcompete a resident population of selfish strains. By investigating 
structured and well-mixed populations we aim to understand whether the existence of 
group structure is a required condition for the maintenance of efficient metabolism pathways.

\bigskip
We showed that in well-mixed populations the efficient strain is only viable in a subset of the parameter space corresponding
 to the region where there is no social conflict. In other words, the efficient strain can only invade a 
population of inefficient strains when it is rare, under the
condition it is already a strong competitor in a pairwise competition with its opponent. 
The lines delimiting the social conflict region and the isocline $P_{fix}=1$ overlap. 
These lines also delimit the region of stability of the two competing metabolic strategies.
Together, these results suggest that a complete mixing of cells creates an unfavorable scenario for the appearance and subsequent 
maintenance of an efficient mode of metabolism. 

\bigskip
On the other hand, in structured populations a more favorable scenario is presented.
Over a broad domain of the parameter space, which goes further beyond the boundaries settled down by the social conflict
and the invasion analysis, the efficient strain is selected for. This result highlights the role of structuring, and more specifically 
kin selection, in promoting the efficient mode of metabolism. In a population which assumes complete mixing of individuals, 
relatedness is expected to be low, while in structured populations relatedness is expected to be relatively higher. This is corroborated
when we studied the effect of migration on the dynamics of fixation of the efficient strain. As migration is 
reduced, isolation increases, and so does the strength of kin selection. An abrupt soaring of the fixation probability of the
efficient strain was observed. Although local competition among relatives may increase in 
structured populations \shortcite{TaylorEvolEco1992,TaylorEvolEco1992,WestScience2002},
it does not mitigate the benefits of kin-selection in the model. More evidence of this finding is obtained  by checking the
dependence of the fixation probability on the carrying capacity $P_{max}$, whereby it is demonstrated that 
the efficient strain faces an even more favorable environment when smaller
groups are assumed.

\bigskip
So one question remains: which mechanism relieves the effect of increased competition 
in small groups thus turning kin-selection an effective evolutionary force? 
Local competition for resource is attenuated in the structured populations as groups are elastic and
their sizes can vary between fission events. This is consistent with previous findings \shortcite{WestScience2002,KummerliEvolution2009}.
Although a larger group means less resource available per individual, there is
no competition for space. Once one individual reproduces it does not necessarily lead to 
the removal of any interactant.
Additionally, a group containing more efficient individuals has as advantage, as a larger amount 
of resource will be available and the net amount of energy converted in the process can be larger. This 
allows the group to reach the carrying
 capacity $P_{max}$ faster. The fact that overlapping generations increase relatedness also contributes to enhance the
chance of the efficient trait \shortcite{WestScience2002}. The third mechanism is the process of group split, which 
produces a two-fold benefit: first 
by reducing local competition because after division the number of cells 
per group drops; second, if the number of efficient cells that occupy a given group
increases, so does the chance that a new-born group is only filled by efficient strains, thus helping to perpetuate
the trait. In summary, three premises of the model ---  group expansion,
overlapping generations and group split --- contribute to attenuate local competition thus
allowing a net benefit effect of kin selection.

\bigskip
The observation of a critical value
of $P_{\rm{max}}$ imposes a threshold for kin selection to act as an effective force. Larger group sizes entail 
smaller relatedness in such way that Hamilton's rule $bc > r$ is no longer followed. 
The group split is a key premise of the model. In a recent work, Ratcliff et al. designed an experiment 
to understand the evolution of complex multicellular organisms from unicellular
ancestors \shortcite{RatcliffPNAS2012}. In the experiment, a population of yeast ({\it S. cerevisiae}) was 
artificially selected for multicellularity by growing in nutrient-rich medium. Serial transfers of the bottom fraction of the liquid were 
performed, and cells collected together with the liquid formed the next generation. After few transfers, they 
already observed that populations were dominated by
snowflake-shaped multicellular clusters. Their witness of the occurrence of cluster reproduction is a basic premise to
settle down life in its multicellular form. The authors ascertained that the daughter clusters were produced as
multicellular propagules rather than mass dissolution of parental clusters. They also observed the existence of a threshold for the 
minimum size of daughter clusters \shortcite{RatcliffPNAS2012}. In our viewpoint, the findings in the experiment of Ratcliff et al. 
are in line with the assumptions of the group dynamics of the proposed model in the current work. 

\bigskip
As conjectured, the first step in the evolutionary transition from single-celled to multicellular organisms 
most likely involved  the evolution of undifferentiated multicellularity \shortcite{PfeifferPNAS2003}. Our 
results show that it is very unlikely that the efficient mode of ATP production has arisen and has become widespread 
in a scenario of global interaction and competition (complete mixing). Once there are efficient strains arising through mutation, 
the maintenance of those strains in a well-mixed population would be very improbable as they could be easily invaded 
by strains with higher growth rate in spite of its low efficient metabolism.
Alternatively, the appearance and subsequent growth of efficient strains in frequency must come together 
with the emergence of a mechanism that prevents dividing cells from complete separation enabling clusters
through parent-offspring adhesion \shortcite{RatcliffPNAS2012}.

\section*{Acknowledgments}
AA has a fellowship from the Conselho Nacional de Desenvolvimento 
Cient\'ifico e Tecnol\'ogico (CNPq). LF has a fellowship from the Funda\c{c}\~ao de Amparo \`a Ci\^encia  e Tecnologia do Estado de Pernambuco (FACEPE). WH gratefully acknowledges funding from the Max  Planck Society. 
PRAC is partially supported by the Brazilian research
agency CNPq and acknowledges financial support from the Funda\c{c}\~ao de Amparo \`a Ci\^encia  e 
Tecnologia do Estado de Pernambuco (FACEPE). 
FFF is supported by Funda\c{c}\~ao de Amparo \`a Pesquisa do Estado de S\~ao Paulo (FAPESP).

\bibliography{Paper-Submitted}{}
\bibliographystyle{chicago}

\newpage
\renewcommand\thefigure{S\arabic{figure}}
\renewcommand{\theequation}{SI \arabic{equation}}
\setcounter{figure}{0}
\setcounter{equation}{0}
\appendix
{\bf \Large Supporting Information}

\bigskip
\bigskip
\bigskip

\section{The Jacobian matrix}
Here we calculate the eigenvalues of the Jacobian matrix of the discrete model. If the absolute values of the  eigenvalues are smaller than one then
the solution in Section 3.2.1  is stable. If we write the set of equations (10) of our two-variable model as $n_{D}(t+1)=f_{D}(n_{D},n_{C})$ and $n_{C}(t+1)=f_{C}(n_{D},n_{C})$, the Jacobian matrix is defined as
\[ J= \left( \begin{array}{cc}
\frac{\partial f_{D}}{\partial n_{D}} &  \frac{\partial f_{D}}{\partial n_{C}} \\
\frac{\partial f_{C}}{\partial n_{D}} &  \frac{\partial f_{C}}{\partial n_{C}} \end{array} \right).\]
The Jacobian matrix must be evaluated at the equilibrium of interest, i.e. $\hat{n}_{C}=0$ and $\hat{n}_{D}=-\frac{\alpha_{D}^{ATP}S}{\ln (1-\nu/a_{D})}$.
After some algebra, the coefficients of the matrix at the equilibrium of interest are determined 
\[ J= \left( \begin{array}{cc}
1+a_D(1-\nu/a_D)\ln (1-\nu/a_D) &  \frac{a_D}{\epsilon}(1-\nu/a_D)\ln (1-\nu/a_D)\\
0 &  1-\nu+a_{C}\left[1-(1-\nu/a)^{\frac{1}{\Delta_{ATP}\epsilon}}\right] \end{array} \right),\]
where $\Delta_{ATP}=\frac{\alpha_{D}^{ATP}}{\alpha_{C}^{ATP}}$ and $\epsilon=\frac{A_{D}}{A_{C}}$.
The above matrix is upper diagonal which means that the eigenvalues of the Jacobian are simply the diagonal coefficients.  Because
$\nu<a_{D}$ for the referred equilibrium, the element $J_{11}$ is always smaller than one and positive regardless the values 
of $\nu$ and $a_{D}$. Note
that the logarithmic term is always negative. Therefore, the stability of the solution is uniquely settled by the element $J_{22}$.  
Of particular interest is the point at which the equilibrium becomes unstable which occurs when $J_{22}$ equals one. Making $J_{22}$ 
equal to one a relation between $\Gamma_{ATP}$ and $\Delta_{ATP}$ is obtained. 

The same analysis is done in Section 3.2.2, but instead the solution
$\hat{n}_{D}=0$ and $\hat{n}_{C}=-\frac{\alpha_{C}^{ATP}S}{\ln (1-\nu/a_{C} )}$ is considered.

\section{Coexistence solution}
The system of equations (10) also admits a solution where the two strains coexist. By rearranging
the set of equations we verify that the coexistence solution obeys the following set of equations:
\begin{align}\label{family}
& n_{D}\epsilon+n_{C}=-\frac{\alpha_{D}^{ATP}S\epsilon}{\ln (1-\nu/a_{D})} \\
& n_{D}\epsilon+n_{C}=-\frac{\alpha_{C}^{ATP}S}{\ln (1-\nu/a_{C})},
\end{align}
showing that there is an indeterminacy, and the coexistence solution only exists 
when $\Delta_{ATP}=\frac{1}{\epsilon}\frac{\ln (1-\nu/a_{D})}{\ln (1 - \nu/a_{C})}$. Actually Eq. (SI\,1) describes a family of solutions.
Of course, the range of values at which $n_{D}$ and $n_{C}$ are meaningful are  $n_{C} \in [0,\mathcal{N}]$ and
 $n_{D} \in [0,\mathcal{N}]$, where
$\mathcal{N}=-\frac{\alpha_{C}^{ATP}S}{\ln (1-\nu/a_{C})}$ with the 
constraint that $n_{D}\epsilon+n_{C}=-\frac{\alpha_{C}^{ATP}S}{\ln (1-\nu/a_{C})}$. Note that when $n_{C}=0$ we recover the solution 
$n_{D}=-\frac{\alpha_{D}^{ATP}S}{\ln(1-\nu/a_{D})}$, whereas if $n_{D}=0$ we recover the solution 
$n_{C}=-\frac{\alpha_{C}^{ATP}S}{\ln(1-\nu/a_{C})}$. 
By determining the Jacobian of the system at equilibrium and hence calculating
its corresponding eigenvalues we check whether the coexistence solution is stable or not. With the help of {\it mathematica}, we
performed a numerical computation of the eigenvalues for different set of parameters 
by varying all the range of acceptable values of
$n_{D}$ (since $n_{C}$ is written as a function of $n_{D}$). As expected in this situation, one 
of the eigenvalues is equal to one, which
means that any disturbance in the same direction as the line drives the system to the 
new values of $(n_{D},n_{C})$. This is called marginal stability as found in a simple
predator-prey model studied by Lotka and Volterra \shortcite{Parker2010}. The stability
of the family of solutions will be dictated by the second eigenvalue. For the set of parameter values used in our simulations
the second eigenvalue is slightly smaller than one, meaning that any transverse disturbance is damped by 
the system, thus warranting stability of the family of solution in Eq. (\ref{family}). It is important to highlight that
the coexistence solution only holds for $\Delta_{ATP}=\frac{1}{\epsilon}\frac{\ln (1-\nu/a_{D})}{\ln (1 - \nu/a_{C})}$, which
describes a line in the diagram $\Delta_{ATP}$ versus $\Gamma_{ATP}$. 

In addition, from Eq. (10) we also notice that $\hat{n}_{D}=0$ and $\hat{n}_{C}=0$ are also equilibrium solutions, where both types go extinct. The solution is only stable when $\nu > a_{D}$ and  $\nu > a_{C}$.

The coexistence of the two strains can not be observed in the simulations as the family of solutions embodies the absorbing states
$n_{C}=0$ and $n_{D}=0$. As stochasticity plays a key role in finite populations, in a finite time the system will eventually reach
one of these boundaries. Since we always start from $n_{C}=1$, the system will probably evolve to $n_{C}=0$ in a short time.

\section{Population size at equilibrium}
Here we show how the stationary population size
of individuals type $D$ (before introducing a single strain $C$), $N_{\rm{st}}$, depends on the set of parameters. From 
this we gain some insight about the strength of stochasticity.
As intuitively expected,  
a linear relation between population size and the influx of resource into the system is
observed (data not shown). The linearity is not followed when studying the dependence of the population size  on the other parameters.

In Figure S1 we show the population size in terms of
$\Gamma_{ATP}=A^{ATP}_{D}/A^{ATP}_{C}$ and $\Delta_{ATP}=\alpha^{ATP}_D/\alpha^{ATP}_C$, for both structured (left panel) and
panmictic populations (middle panel).  
The values of $\Gamma_{ATP}$ and $\Delta_{ATP}$ are varied by keeping $A^{ATP}_{C}=2$ and $\alpha^{ATP}_C=0.5$ 
and changing $A^{ATP}_{D}$ and $\alpha^{ATP}_D$, respectively. The range
of values of $\Gamma_{ATP}$ and $\Delta_{ATP}$ are, as argued before, those that are meaningful to the problem here addressed. 
Please note that a logarithmic scale is adopted for the color gradation in Fig. S1. The range of
 $\Delta_{ATP}=\alpha^{ATP}_D/\alpha^{ATP}_C$ in which population size is shown for the well-mixed populations is much smaller
than the range displayed for structured population. Outside of this range  the population already falls 
in the regime where the efficient strain is counter-selected - as can be seen in Fig. 2 (Main Text). 
The right panel of Fig. S1
shows the equilibrium solution for the discrete-time model of a well-mixed population, 
given by $\hat{n}=-\frac{\alpha_{D}^{ATP}S}{\ln (1-\nu\frac{E_{max}}{A_D^{ATP}})}$, as derived in Section 3.1.1.
The grey region denotes the extinction of the population, where the above solution becomes unstable, while
the solution $\hat{n}=0$ becomes stable. The plot evinces a very good agreement between the theoretical 
prediction and simulations. The simulation results for a well-mixed population display a larger grey area in comparison to
the theoretical prediction, which owes to stochastic effects, which are not captured by the time-discrete model. 
In the right panel, the population size is too small --- around five or less individuals --- being prone to extinction.

Especially, in structured populations
the stationary population size $N_{\rm{st}}$ of a pure population of strain $D$ can vary from less than $100$ 
individuals (small $\Gamma_{ATP}$ and $\Delta_{ATP}$) to nearly $5,000$ individuals (large $\Gamma_{ATP}$ and 
 $\Delta_{ATP}$). The wide range of the population size at equilibrium
explains why we should adopt a relative measurement of fixation probability since the strength of drift, namely $1/N_{\rm{st}}$, is variable
under the change of the parameter values.

\begin{figure}[htp]
\centering
\includegraphics[width=\textwidth]{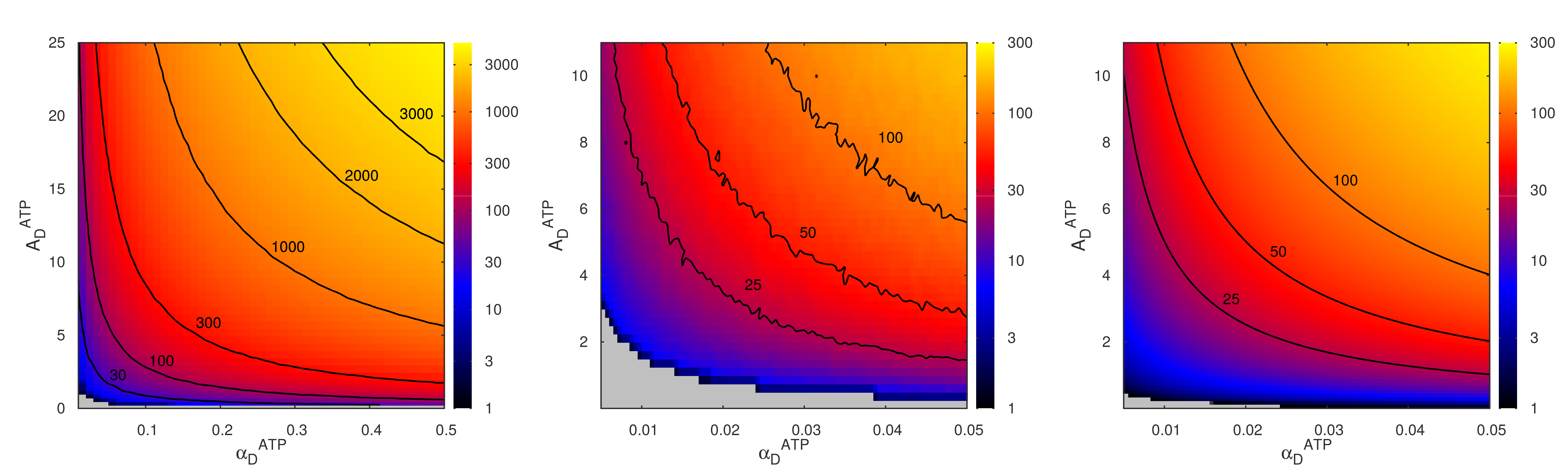}
\caption{Stationary population size of defectors. The figure is a heat map of the population size of defectors (just before the introduction of a single cooperator) in terms of $A^{ATP}_D$ and $\alpha^{ATP}_D$. Left panel: simulation results for structured populations, middle panel: simulation results for well-mixed populations and right panel: theoretical prediction, $n_{D}=-\frac{\alpha_{D}^{ATP}S}{\ln (1-\nu E_{\rm{max}}/A_{D}^{ATP})}$. The other parameter values are influx rate of resource $S=25$, death rate $\nu=0.01$, group carrying capacity $P_{\rm{max}}=10$, internal energy for one individual to split to two individuals $E_{\rm{max}}=10$, and $A_{D}=10$. The simulation data points correspond to $40$ distinct populations and for each population $10,000$ independent runs were performed. The black lines denote isoclines, along which the population size is a constant (corresponding values indicated). The grey region denotes a region where the defector population is not sustainable and goes to extinction (before inserting the cooperator).}
\label{fig:S1}
\end{figure}

\end{document}